\RequirePackage{fix-cm}
\documentclass[smallextended]{svjour3}       
\smartqed  
\usepackage{graphicx}
\usepackage{booktabs} 
\usepackage{array,multirow}
\usepackage[ruled,noend]{algorithm2e} 
\usepackage{url} 
\usepackage{natbib}
\usepackage{amssymb}
\usepackage{lscape}
\usepackage{mathrsfs}
\usepackage{subfig}

\begin{document}

\title{A Lightweight Decentralized Service Placement Policy for Performance Optimization in Fog Computing\thanks{This research was supported by the Spanish Government (Agencia Estatal de Investigaci\'on) and the European Commission (Fondo Europeo de Desarrollo Regional) through grant number TIN2017-88547-P (MINECO/AEI/FEDER, UE).}
}

\titlerunning{A Lightweight Decentralized Service Placement Policy}        

\author{Carlos Guerrero         \and
	Isaac Lera \and Carlos Juiz 
}


\institute{C. Guerrero (Corresponding author), I. Lera, C. Juiz \at
	Computer Science Department, University of Balearic Islands, Crta. Valldemossa km 7.5, E07122 Palma, Spain \\
	Tel.: +34-971172965\\
	Fax: +34-971173003\\
	C. Guerrero\\
	\email{carlos.guerrero@uib.es} \\
	I. Lera\\
	\email{isaac.lera@uib.es} \\
	C. Juiz\\
	\email{cjuiz@uib.es}           
}

\date{Received: date / Accepted: date}

\maketitle

\begin{abstract}
	A decentralized optimization policy for service placement in fog computing is presented. The optimization is addressed to place most popular services as closer to the users as possible. The experimental validation is done in the iFogSim simulator and by comparing our algorithm with the simulator's built-in policy. The simulation is characterized by modeling a microservice-based application for different experiment sizes. Results showed that our decentralized algorithm places  most popular services closer to users, improving network usage and service latency of the most requested applications, at the expense of a latency increment for the less requested services and a greater number of service migrations.

\end{abstract}

\section{Introduction}\label{sec1}

The emerging of application development for Internet of Things (IoT) has revitalized the popularity of wearables, logistics, smart cities, or e-health environments \citep{ATZORI20102787,Li:2015:ITS:2750609.2750625,Ko:2016:STU:2909066.2835492,Darwish2017a}. The number of users in these systems, and their performance requirements, is increasing continuously.  Those requirement increments were initially satisfied by integrating IoT and cloud architectures~\citep{BOTTA2016684,DIAZ201699}. The combination of both technologies allows, firstly, IoT applications to dispose of unlimited computational and storage capacities, and secondly, to expand the scope of cloud systems by dealing with real-life components~\citep{CAVALCANTE201617,Darwish2017b}. But the integration of IoT and cloud generates new problems, as for example the increase in the service latency that is a critical requirement for some e-health or gaming IoT applications~\citep{7543455}. Fog computing emerged to cover these limitations and it opened a broad range of renewed challenges in topics such as security, reliability, sustainability, scaling, IoT marketplaces, or resource management~\citep{VARGHESE2017,Mahmud2018,7498684}. 

Fog computing exploits resource capacities of networking components to allocate services or to store data.  Therefore, firstly, services are closer to the clients and, secondly, data do not need to be transfered in its whole to the cloud. Both issues have a notable performance impact, reducing the network latency and usage. Resource managers are important components to improve this performance.

On the one hand, fog data managers need to deal with the selection of the data that is stored in the cloud providers and which is stored in fog devices. For the latter case, placement policies need to be considered to optimize the use of the storage capacities of the fog layer. On the other hand, fog service orchestrators need to decide the allocation of the services in the fog devices to improve the Quality of Service (QoS) and the usage of the fog layer considering scalability and dynamicity requirements~\citep{7867735}. They should deal with the placement and scalability of user-shared services by creating, in the fog devices, additional instances of services already available in the cloud.


IoT environments, as for example smart cities, usually have thousands, even millions, of devices, clients or services. Centralized management could reach in a non-affordable optimization. Previous solutions to solve the Fog Service Placement Problem (FSPP) are based on centralized solutions, with the following main drawbacks: (a) scalability, as the execution time of a global optimization algorithm is usually increased as the number of devices to manage is increased; (b) network overhead, as the fog devices need to send their performance data to the centralized broker; (c) reliability, as the broker is a single point of failure (SPOF); (d) decision latency, as the decisions need to be transmitted from the centralized broker to each device; and (e) heterogeneity, as the central broker needs to deal with data from very different types of fog devices, and the communication protocol or the orchestrating mechanism could be also very different. 

We propose to address the problem of service placement in a decentralized way, where each device takes the local optimization decisions, by considering only its own resource and usage data and consequently, the execution time is independent of the number of devices, the performance data is not send between devices, SPOF are avoid, the decision are taken locally and all the resources and the inputs of the algorithm are homogeneous inside of the same device. We propose to base the decisions of the algorithm in placing the most popular services as closer as possible to the clients, using the hop distance as the indicator of this proximity. 

We raise two research questions: (RQ1) Is it possible to define a local and decentralized optimization algorithm that would be able to place the most popular services in the fog devices with smallest hop distances to the clients?; (RQ2) Would this placement policy result in an optimization of the performance of the fog computing architecture in terms of, for example, service latency or network usage?


The three main contributions of our work are: (a) An up-to-date brief survey of researches addressed to optimize the Fog Service Placement Problem (FSPP); (b) A decentralized and low overhead proposal to reduce the network usage in a fog computing architecture; (c) An experimental validation based on a microservice-based application.

\section{Related Work}
\label{relatedsect}


Previous fog service management algorithms have explored a wide range of optimization techniques, such as heuristics, greedy algorithms, linear programming, or genetic algorithms, between others. These service managers have defined several aspects of the fog resources, such as placement, scheduling, allocation, provisioning, or mapping for services, resources, clients, tasks, virtual machines, or even fog colonies.  These solutions have been defined for environments such as industrial IoT, smart cities, eHealth, or mobile micro-cloud.


The characteristics of the related work have been summarized in Table~\ref{survey1}, by indicating the IoT scope for which the solution was proposed (column \textit{Scope}), the optimization purpose (column \textit{Objective functions}), the elements that the optimization algorithm manages to improve the objective functions (column \textit{Decision variables}), if the broker manager is centralized or not (column \textit{Broker}), the optimization algorithm (column \textit{Alg.}), and the technique or tool used in the validation (column \textit{Val.}). The related studies are grouped in terms of their optimization algorithms.

We explain the related researches grouped by their optimization algorithms.

\begin{table}%
	\caption{Summary of the brief survey for Fog Service Placement Problem approaches.}
	\label{survey1}
	\scriptsize
	
	\begin{minipage}{\columnwidth}
		\begin{center}
			\begin{tabular}{p{0.1\columnwidth}p{0.05\columnwidth}p{0.25\columnwidth}p{0.25\columnwidth}p{0.05\columnwidth}p{0.06\columnwidth}p{0.06\columnwidth}}		
					\hline\noalign{\smallskip}
				\textbf{Authors} &\textbf{Scope}\footnote{\emph{Scope:} G- General IoT systems; I - Industrial IoT systems; H - eHealth IoT systems; C - Crowdsensing apps; E- Embedded systems; M - Mobile Micro-Clouds (MMC).}& \textbf{Objective functions}  & \textbf{Decision variables} &  
				\textbf{Broker}\footnote{\emph{Broker:} C - Centralized; D - Decentralized; Ds - Distributed ; H - Hierarchical (Two broker levels, global and fog colonies).}&  \textbf{Alg.}\footnote{\emph{Algorithm:} ILP - Integer Linear Programming; MILP - Mixed integer linear programming; GA - Genetic Algorithm; MDP - Markov Decision Process; D-MDP - Decoupled Markov Decision Process; MWIS - Maximum Weighted Independent Set; ADMM - Alternating Direction Method of Multipliers; VM - Virtual Machine; PN - Petri Nets; MC - Monte Carlo; Cons. - Consensus algorithm;  SP - Shortest-path; Heur. - Heuristics; BP - Binary programming; D - Decomposition; BET - Benchmarking, Evaluation and Testing; Own - Own algorithm.}&\textbf{Val.}  \footnote{\emph{Validation:} O - Own simulation; F - iFogSim; A - Analytically; B - Benchmarking;  P - PuLP and Gurobi Optimizer; T - Testbed.}\\
	\noalign{\smallskip}\hline\noalign{\smallskip}
				Arkian~\citeyear{ARKIAN2017152} & C &Cost  & Client association, Resource provisioning , Task distribution, VM placement& C & MILP & O\\
				
				Gu~\citeyear{7359164} & H & Cost & Base station association, Task distribution, VM placement & C & ILP & O\\
				
				Velas\-quez~\citeyear{Velasquez2017} & G & Network latency, Service migrations & Service placement & C & ILP & ---\\
				
				Huang~\citeyear{HUANG201447,Huang2014}& G & Communication power consumption & Service merging and placement& C  & ILP, MWIS & O\\
				
				Souza~\citeyear{7511465} & G & Service delay & Service placement & C & ILP & P\\
				
				Skarlat~\citeyear{8014364} & G& Deadline violations, Cost, Response time& Service placement & C  & ILP & F\\
				
				Zeng~\citeyear{7422054}& E & Task completion time & Task scheduling, Task image placement, Workload balancing & D & MILP & O\\
				
				Barcelo~\citeyear{7676307} & G & Power consumption & Service placement & C & ILP & O\\
				
				Wen~\citeyear{7867735} & G & Response time, QoS & Service placement & C  & GA & O\\
				
				Skarlat~\citeyear{Skarlat2017} & F & Resource waste, Execution times & Fog colony service placement & H  & GA & F\\
				
				Yang~\citeyear{7110527} & M & Cost, Latency and Migration & Service placement and Load dispatching & C & Greedy, ILP, GA & O\\
				
				Ni~\citeyear{7935527} & G & Response time, Cost & Resource allocation and scheduling  & C   & PN &O\\
				
				Urga\-onkar~\citeyear{URGAONKAR2015205} & G & Queue length, cost & Service migration & C &D-MDP& O\\

				Brogi~\citeyear{7919155, 8014366} & G & Resource consumption, QoS & Service placement & C & MC & O\\

				Colistra~\citeyear{COLISTRA201498} & G & Resource usage& Resource allocation & Ds & Cons. & O \\
				
				Wang~\citeyear{7249199,Urgaonkar:2015:DSM:2822545.2822799} & M & Cost & Look-ahead service placement & C & SP and MDP & O \\
				
				Billet~\citeyear{7035717} & G & Resource usage, Power consumption and Load balancing & Task placement & C & BP Heur. Greedy& O \\ 
				
				Taneja~\citeyear{7987464}  & G & Network usage, Power consumption, Latency & Service placement & C & Own & F\\
				
				Wang~\citeyear{7847322} & M & Load balancing & Service placement & C & Own & O \\
				
				Bitten\-court~\citeyear{7912261} & G & Network usage and delay & Service placement & C & Own & F \\
				
				Farris~\citeyear{7389120} & F & Executed tasks & Resource provisioning& C & Own &O\\

				Deng~\citeyear{7248934} & G & Power consumption, Response time & Workload placement & C & D & \\

				Saurez~\citeyear{Saurez2016} & G & Latency & Service placement and migration & C & Own&B  \\
				
			    Venti\-cinque~\citeyear{Venticinque2018}& G & Performance & Service placement & C & BET & T \\
				
				Gupta~\citeyear{SPE:SPE2509}& G & Energy, Network usage, Latency & Sevice placement & D & FIFO & F\\

				\\
				
				[This work] & G & Hop count & Service placement & D & Own & F\\

		\noalign{\smallskip}\hline
			\end{tabular}
		\end{center}
		\bigskip\centering
		\footnotesize

	\end{minipage}
\end{table}%

Linear programming is a common approach for resource optimization. \citet{ARKIAN2017152} formulated a mixed-integer non-linear program that was linearized into a mixed integer linear program for the optimization of cost. \citet{7359164} also used this optimization approach. They integrated medical cyber-physical systems and fog computing and optimized the cost by considering the base station association, task distribution and virtual machine placement.  The work of \citet{Velasquez2017} was addressed to reduce the number of service migrations and the network latency.

\citet{HUANG201447} presented a quadratic programming formulation for the problem of reducing the power consumption in fog architectures by co-locating neighboring services on the same devices. \citep{Huang2014} also presented a previous work where the problem was modeled as a Maximum Weigthed Independen Set problem (MWIS). \citet{7511465} studied an allocation algorithm based on Integer Linear Programming that minimized the service latencies in a fog computing environment, while the fulfillment of capacity requirements were guaranteed. \citet{8014364} studied the service placement problem by considering the QoS requirements of the applications executed in a fog architecture. \citet{7422054} proposed to manage the task scheduling, the task storage placement and the I/O balanced use to reduce the task completion time in software-defined embedded systems. \citet{7676307} formulated a service placement optimization to reduce power consumption in IoT environments as a minimum mixed-cost flow problem.

A second set of studies were implemented with genetic algorithms (GA). \citet{7867735} presented a parallel GA to reduce the response time. \citet{Skarlat2017} introduced the concept of fog colonies for a hierarchical optimization process. Each colony used a GA to decide the services that were placed in the colonies and which ones were propagated to neighbor colonies. \citet{7110527} compared three optimization algorithms based on a greedy heuristic, a linear programming and a GA. Additionally, a model to predict the distribution of user's future requests was also presented to adapt the service location.

Linear programming and GA are the most popular solutions, but there are also an important number of related researches exploring other alternatives, such as, Petri nets, Markov decision or new and own algorithms.  \citet{7935527} proposed to use priced timed Petri nets (PTPNs) for resource allocation in fog computing. Their proposal optimized price and time costs to complete a task. Their simulation results showed a higher efficiency than static allocation strategies. \citet{URGAONKAR2015205} addressed the objective of minimizing the operational costs of service placement in fog, while the performance was guaranteed. They modeled the scenario as a sequential decision using a Markov Decision Process (MDP), decoupling de problem into two independent MDPs and, finally, optimizing with the Lyapunov technique. Brogi et al. presented the tool FogTorch in two research works~\citep{7919155,8014366}. The first one presented the model for the QoS-aware deployment of multicomponent IoT applications in fog infrastructures. The second one presented the results of using Monte Carlo simulations in the FogTorch tool, classifying the deployments in terms of QoS and resource consumptions. \citet{7057905} proposed a decentralized algorithm for resource allocation in fog environments for the specific case of video streaming. Their solution was based on the proximal algorithm and alternating direction method of multipliers. The model was validated analytically.  \citet{COLISTRA201498} adapted the consensus algorithm to allow devices to cooperate in the distributed resource allocation problem to adequately share the resources. \citet{7249199} presented a service placement algorithm that used predicted costs for look-ahead optimizing the total provider cost for a given period of time. The authors validated their proposal using real-world user-mobility traces in simulations. \citet{Urgaonkar:2015:DSM:2822545.2822799} also proposed a solution for this last scenario but considering a Markov Decision Process. \citet{7035717} formulated the task allocation problem in IoT as a binary linear optimization that its computational cost was reduced by including a heuristic and a greedy algorithm.  

\citet{7935527} proposed to use priced timed Petri nets (PTPNs) for resource allocation in fog computing. Their proposal optimized price and time costs to complete a task. \citet{URGAONKAR2015205} addressed the objective of minimizing the operational costs of service placement in fog, while the performance was guaranteed. They modeled the scenario as a sequential decision using a Markov Decision Process (MDP), decoupling the problem into two independent MDPs and, finally, optimizing with the Lyapunov technique. Brogi et al. presented the tool FogTorch in two research works (~\citep{7919155,8014366}). The first one presented the model for the QoS-aware deployment of multicomponent IoT applications in fog infrastructures. The second one presented the results of using Monte Carlo simulations in the FogTorch tool, classifying the deployments in terms of QoS and resource consumptions.  \citet{COLISTRA201498} adapted the consensus algorithm to allow devices to cooperate in the distributed resource allocation problem to adequately share the resources. \citet{7249199} presented a service placement algorithm that used predicted costs for look-ahead optimizing the total provider cost for a given period of time. The authors validated their proposal using real-world user-mobility traces in simulations. \citet{Urgaonkar:2015:DSM:2822545.2822799} also proposed a solution for this last scenario but considering a Markov Decision Process. \citet{7035717} formulated the task allocation problem in IoT as a binary linear optimization that its computational cost was reduced by including a heuristic and a greedy algorithm.  

\citet{7987464} proposed a service placement algorithm for efficient use of the network and power consumption. The algorithm sequentially assigned the highest demanding application modules to the nodes with biggest capacities. \citet{7847322} proposed, a first optimization stage modeled as linear graphs to be later extended to tree application models using algorithms with polynomial-logarithmic ratios. \citet{7912261} compared three service allocation algorithms to illustrate that these strategies depend on the demand coming from mobile users and can take advantages  of fog proximity and cloud elasticity.  \citet{7389120} proposed that the edge nodes orchestrated  the provisioning of resources in micro-cloud federations using a decomposer module. Deng et al.~\citeyear{7248934} studied the trade-off between power consumption and delay in fog computing. They decomposed the initial allocation problem into three subproblems independently solved with convex optimization, ILP and Hungarian method. \citet{Saurez2016} presented a service migration algorithm based on the mobility pattern of the sensors and the dynamic computational needs of the applications. The solution was built using containers and the experimental results showed improvements in the migration latencies. \citet{Venticinque2018} presented a methodology based on three phases: Benchmarking, Evaluation and Testing. This methodology helps developers to meet application requirements and to optimize performance and utilization of available resources.

The developers of the iFogSim simulator implemented a decentralized service placement policy called Edgewards~\citep{SPE:SPE2509}. It placed the services in each single path between clients and the cloud. The services were placed in a First-In-First-Allocated policy. Services from different paths were merged if they were placed in the same device and migrated to upper devices if necessary. Additionally, instances from other paths in upper devices of a candidate service were considered and it was placed in the upper device to merge both instances even when closer devices had enough resources. Despite the placement algorithm was decentralized, it needed some general information of the placement status, such as the already placed services for each path between the clients and the cloud. Their experiments compared the results with the allocation of all the services in the cloud provider.

Most of those previous works were modeled as a centralized broker or orchestrator that needs information from all the components in the system (fog devices, clients, cloud, services) and takes global decisions to optimize the service placement. Problems with the scalability and the computational complexity of the algorithm are clear when the number of elements is very high such as, for example, in smart cities.  Decentralized service orchestration in fog computing arises as a current open challenge. It is necessary to define solutions that deal with a smaller number of elements, as in the case of fog colonies~\citep{Skarlat2017}, or even completely decentralized optimizations. We propose a decentralized service placement orchestrator to minimize the hop count of the most requested services.

\section{Architecture Proposal}
\label{architectureproposal}

Fog computing is an architecture pattern where clients request services to cloud providers through a network composed by fog devices. These devices have computational and storage capacities that allow them to allocate data and instances of the cloud services. Therefore, data and service management policies are needed to decide when and where to place the services and the data. Our architecture proposal is focused on the Fog Service Placement Problem (FSPP). 

\begin{figure}
	\includegraphics[width=240pt]{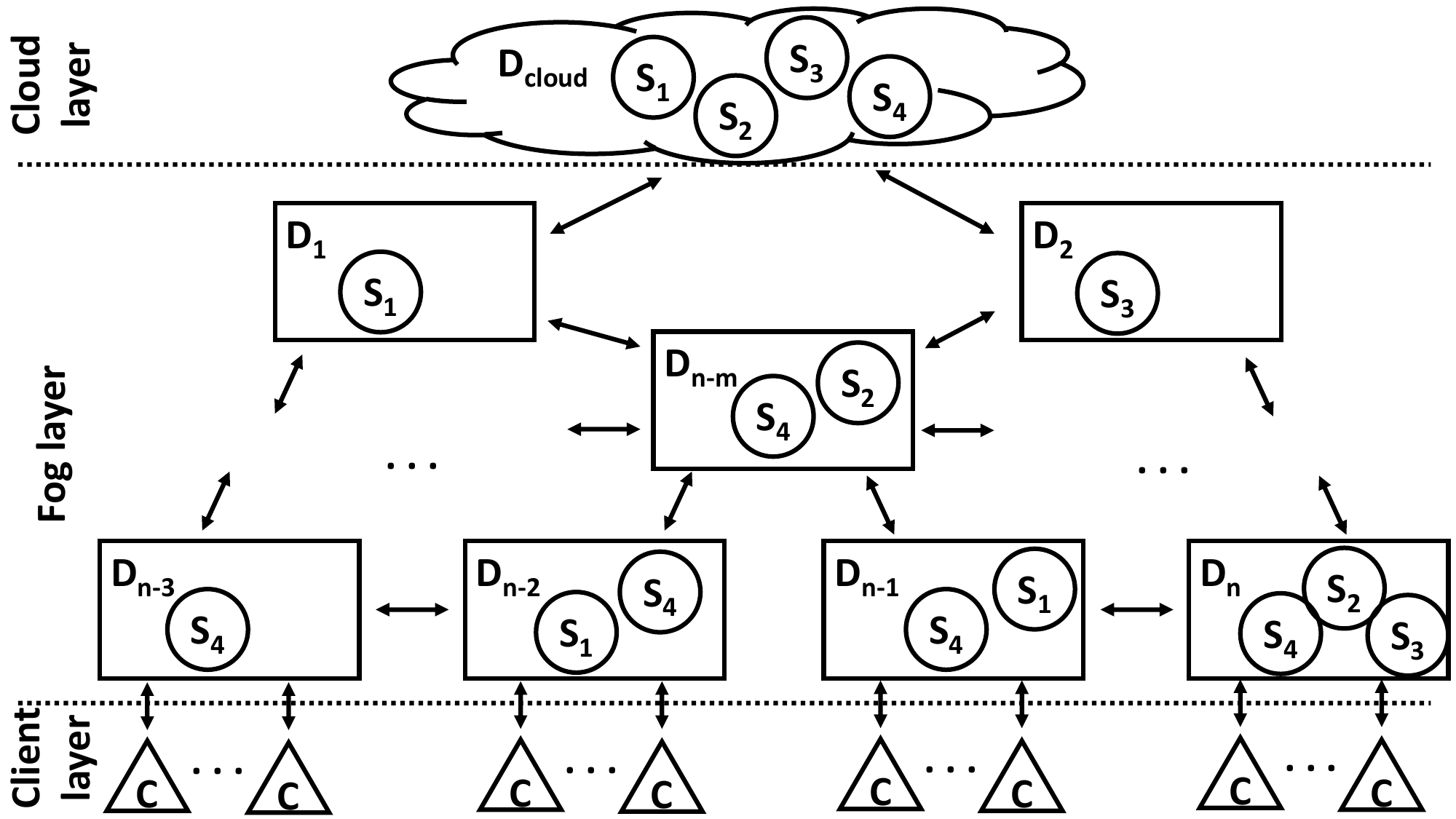}
	\caption{Fog computing architecture.}
	\label{fogarchitecture}
\end{figure}

A general fog computing architecture is represented in Figure~\ref{fogarchitecture} where three layers can be identified: cloud layer, fog layer and client layer. The architecture can be modeled as a graph where the nodes are the devices and the edges the direct network links between devices. Three types of devices can be differentiated: a device for the cloud provider of the cloud layer; the gateways, that are the access points for the clients; the fog devices, the network devices between the cloud provider and the gateways. All the devices have resources to allocate and execute services. 

We consider that the applications follows a microservice based development pattern, that is increasingly being used in IoT applications~\citep{Vogler:2016:SFP:2909066.2850416,7300793,Saurez2016}. This type of applications are modeled as a set of small and stateless services that interoperate between them to accomplish a complex task~\citep{7436659}. Thus, the services can be easily scale up, by downloading the code and executing it, or scale down, by just stopping the execution of the service. We assume that there is at least one instance of each service running in the cloud provider.  


We base our proposal on the idea that the best placement of the services is in devices as closer as possible to the clients' gateways. In an unreal scenario with unlimited resources in the fog devices, the optimal service placement would be to place instances of each service in all the gateways. Since the resource capacity limits the number of service instances in the devices, some services need to be migrated to other devices. Our proposal migrates those services to devices within the shortest path between the gateway and the cloud provider. 

\begin{figure}
	\includegraphics[width=240pt]{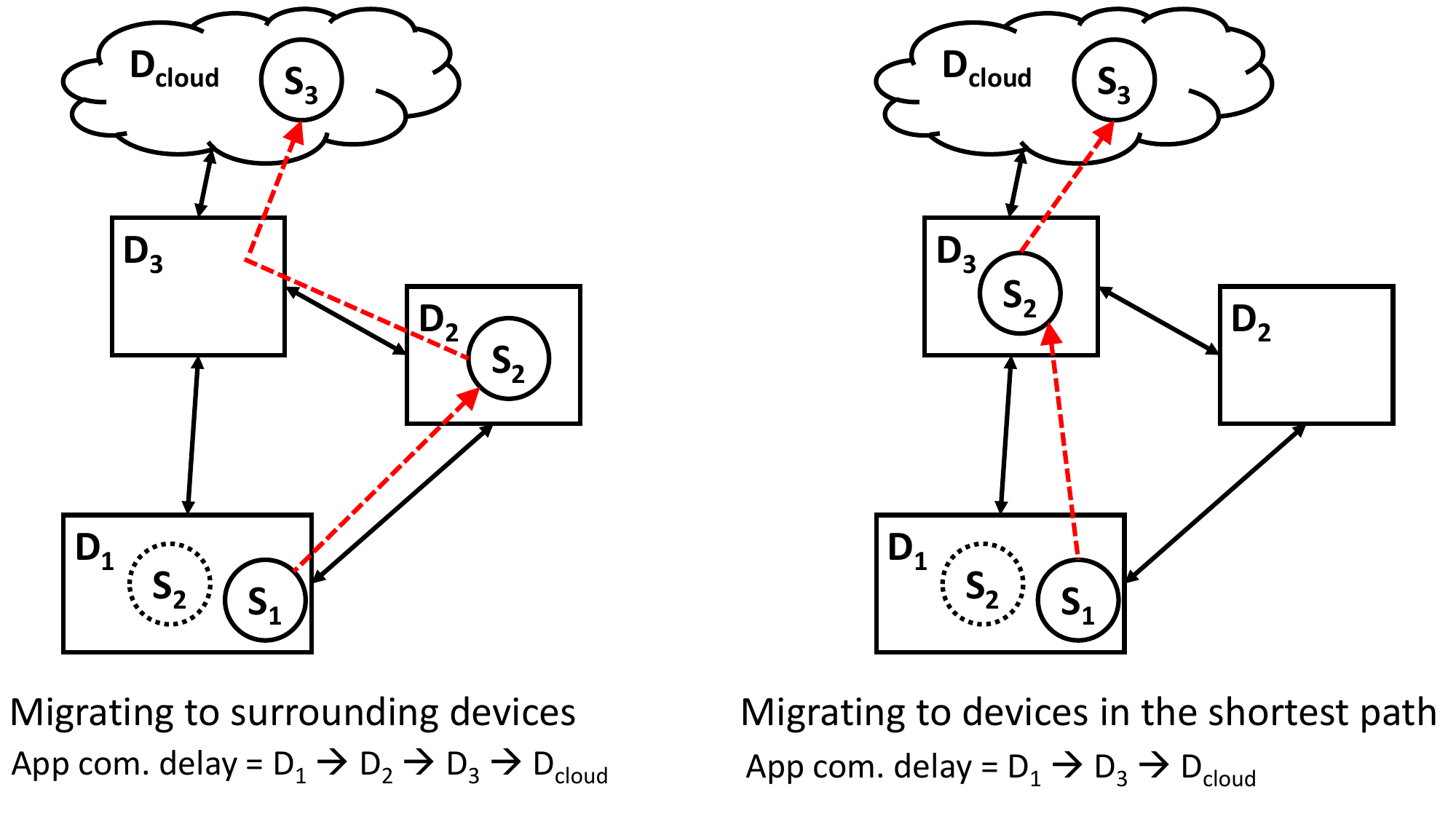}
	\caption{Example of network delay benefits for a service migration scheme within the shortest path.}
	\label{shortestPath}
\end{figure}

Migrating the services along the devices in the shortest path with the cloud provider, instead of any other devices around them, is based on the idea that, sooner or later, the execution flow of the interrelated services would need to execute some service in the cloud provider. Moreover if data need to be stored in a centralized place. Thus, if the migrated services are placed in devices out of the shortest path, the total path to the cloud provider will be increased. Figure~\ref{shortestPath} illustrates an example where a service, $S_2$ is migrated from a device $D_1$. The application communication times is one step bigger when service $S_2$ is placed in $D_2$, a device out of the shortest path between $D_1$ and the cloud provider, with regard to the case of placing it in the next device in the shortest path, $D_3$.

We propose a priority rule by placing first the most popular services, in terms of service requests, in the devices that are closer to the clients, and migrating the less requested to \textit{upper devices} in the shortest path with the cloud provider in case that the closest device does not have enough free resources. For the sake of simplicity, in the rest of the article \textit{upper devices} refers to any of the devices that are in the shortest path between the device and the cloud provider. 

\begin{figure}
	\includegraphics[width=240pt]{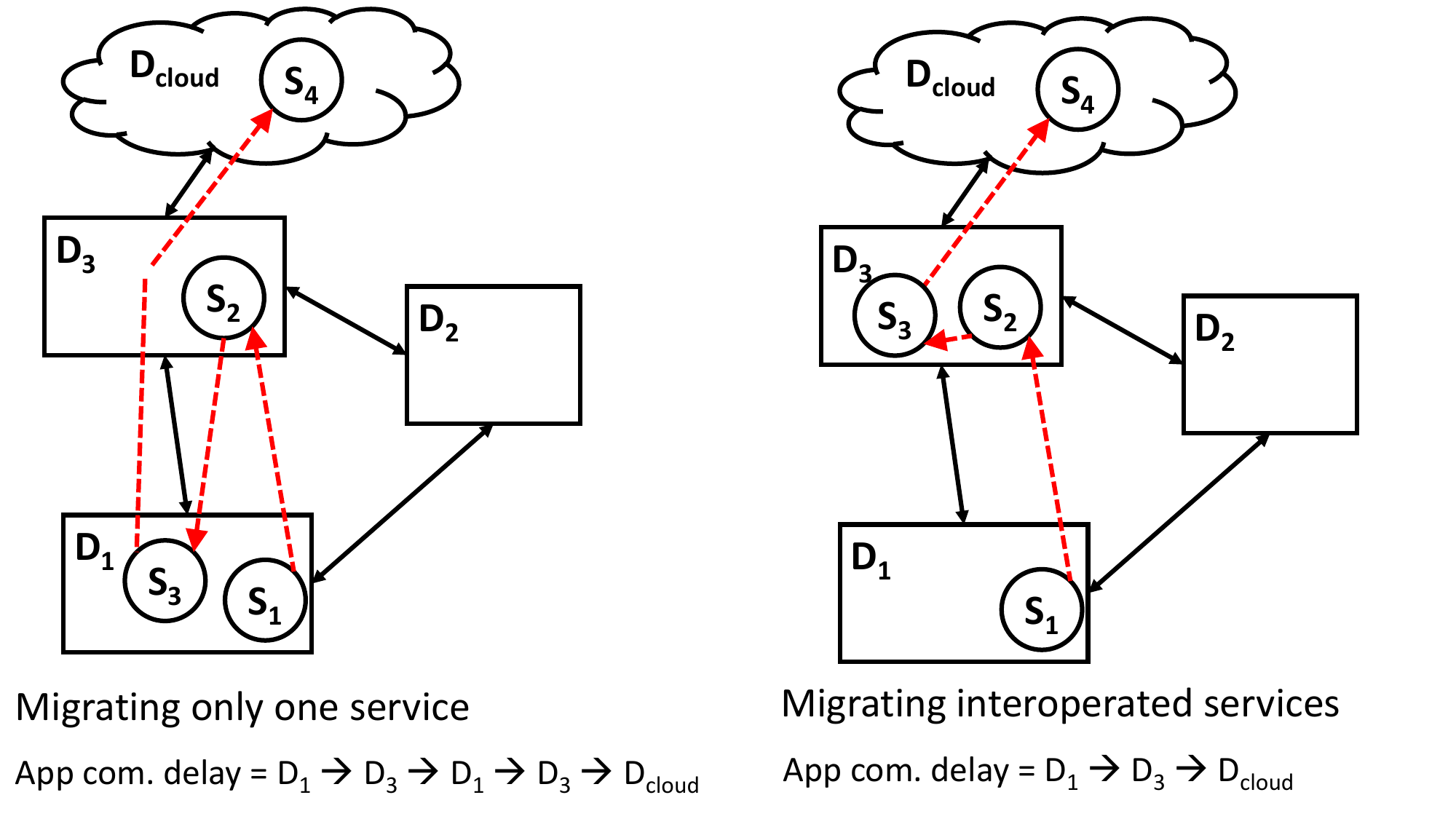}
	\caption{Example of network delay benefits for a service migrating scheme with migration of interoperated services.}
	\label{deviceLoop}
\end{figure}

We finally consider that once that a service is migrated to an upper device, it is also better to migrate all its interoperated services, to avoid device loops in the service execution flow. An illustrative example is showed in Figure~\ref{deviceLoop}. Consider a service execution flow as $S_1 \rightarrow S_2 \rightarrow S_3 \rightarrow S_4$ and that the service $S_2$ needs to be migrated to the upper device $D_3$.  If we only migrate service $S_2$ to device $D_3$, the application makespan will be increased in two communication steps, with regard to the case of also migrating service $S_3$ to $D_3$, which will keep the same application makespan. This strategy is reinforced with the idea that once that a service has been migrated to an upper devices due to not enough free resources, the placement of any of the remaining services will be also very unlikely in the lower device.

\begin{figure}
	\includegraphics[width=240pt]{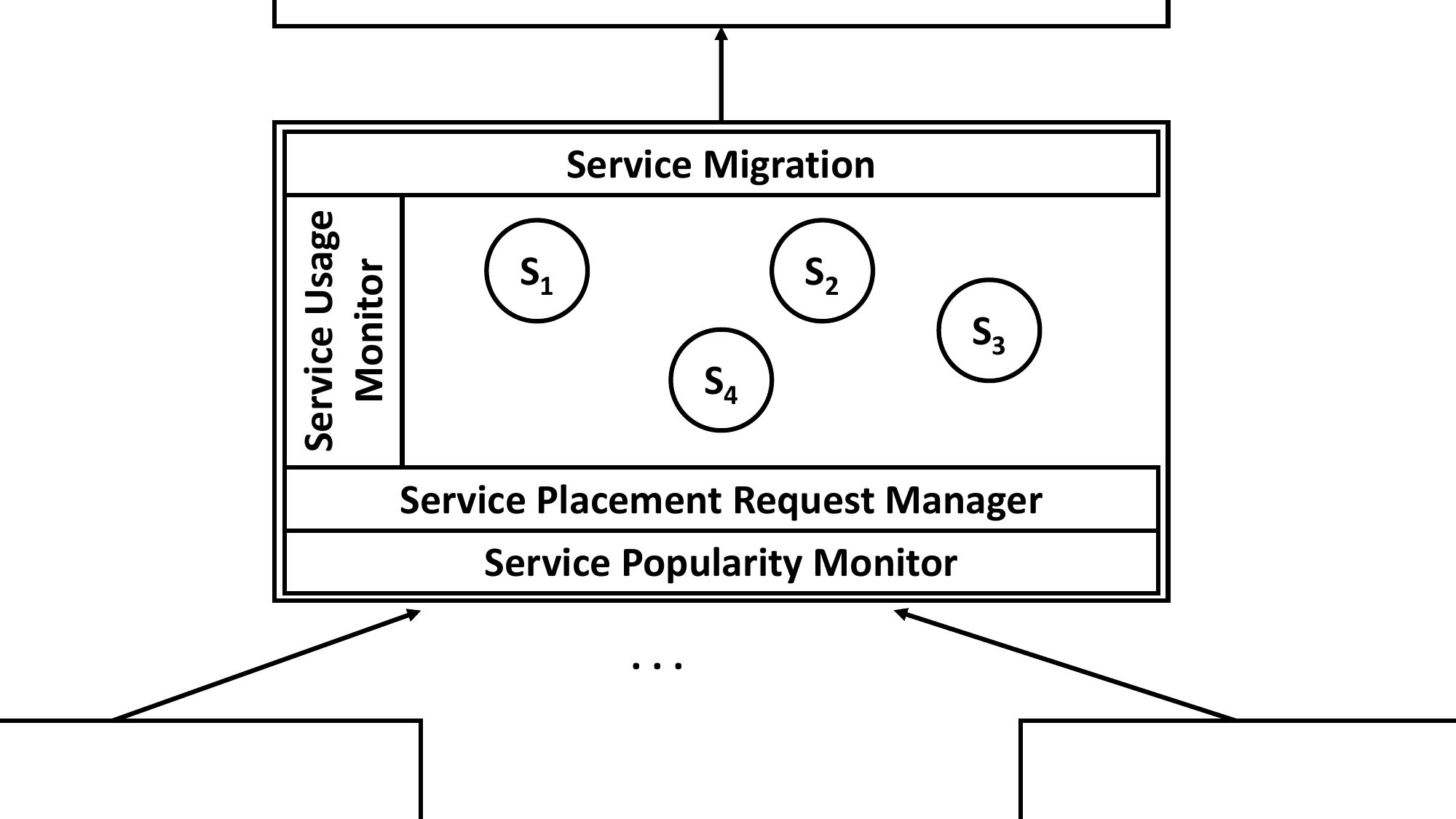}
	\caption{Decentralized service placement manager.}
	\label{devicearchitecture}
\end{figure}

 We propose to execute a decentralized service broker in each device to implement our strategy. Figure~\ref{devicearchitecture} shows the components of the device broker: Service Usage Monitor (SUM), gathers information of the services' performance and resource usage and the information about the services interoperability; Service Migration (SM), sends allocation requests to other devices; Service Placement Request Manager (SPRM), is the local and decentralized optimization algorithm that decides if a given service is allocated or, on the contrary, migrated to other device; and Service Popularity Monitor (SPM), gathers information about the request rate of each service.

When a new client is connected to one leaf device, or gateway, one Service Allocation Request (SAR) is sent to the gateway for each service the client requires. All these requests are received by the SPRM and it decides if the service is placed in the current device or whether the  SAR is shifted to upper devices, based on the algorithm explained in Sect.~\ref{optimizationmodel}. If the SAR is accepted by the SPRM, a service image is downloaded from the cloud provider to be executed in the current device. The placement process could generate deallocation of other services already instantiated in the device. A SAR is sent to upper devices for each deallocated service instance. Deallocated instances are just deleted from the device, as the new instances in the upper devices would be downloaded from the cloud provider. All this process is repeated recursively in the upper devices when a SAR is shifted.

The algorithm is running in each fog device, and the algorithm only considers variables that are obtained locally in the device by the defined monitors (SPM and SUM), such as the device service request rate, or the device resources (demanded, used or available), between others. Sending performance or system data between the devices is not necessary and delays in the decision due to network transmission or overheads in the network are avoid. Additionally, the overall computational complexity of the algorithm is very low, and the placement decision is obtained in a limited period of time and, moreover, this time is not increased as the number of devices is scaled up.


\section{Problem Statement}

\subsection{System Model}

The Fog Service Placement Problem (FSPP) considers a set of clients $C_n$ that request applications that are hosted in a cloud provider $D_{cloud}$. The applications are modeled as a set of services modules, $S_x$ that are related through a many-to-many consumption relationships, $cons: \{S_x\} \rightarrow \{S_{x'}\}$, as the microservice-based application development model defines~\citep{7436659}. This model has been also proposed to deploy applications in fog computing~\citep{Vogler:2016:SFP:2909066.2850416,7300793,Saurez2016}. Applications are defined as a directed graph, where the nodes are the services and the edges indicate the services that are requested by other ones, i.e. their interoperability. 

\begin{figure}
	\includegraphics[width=240pt]{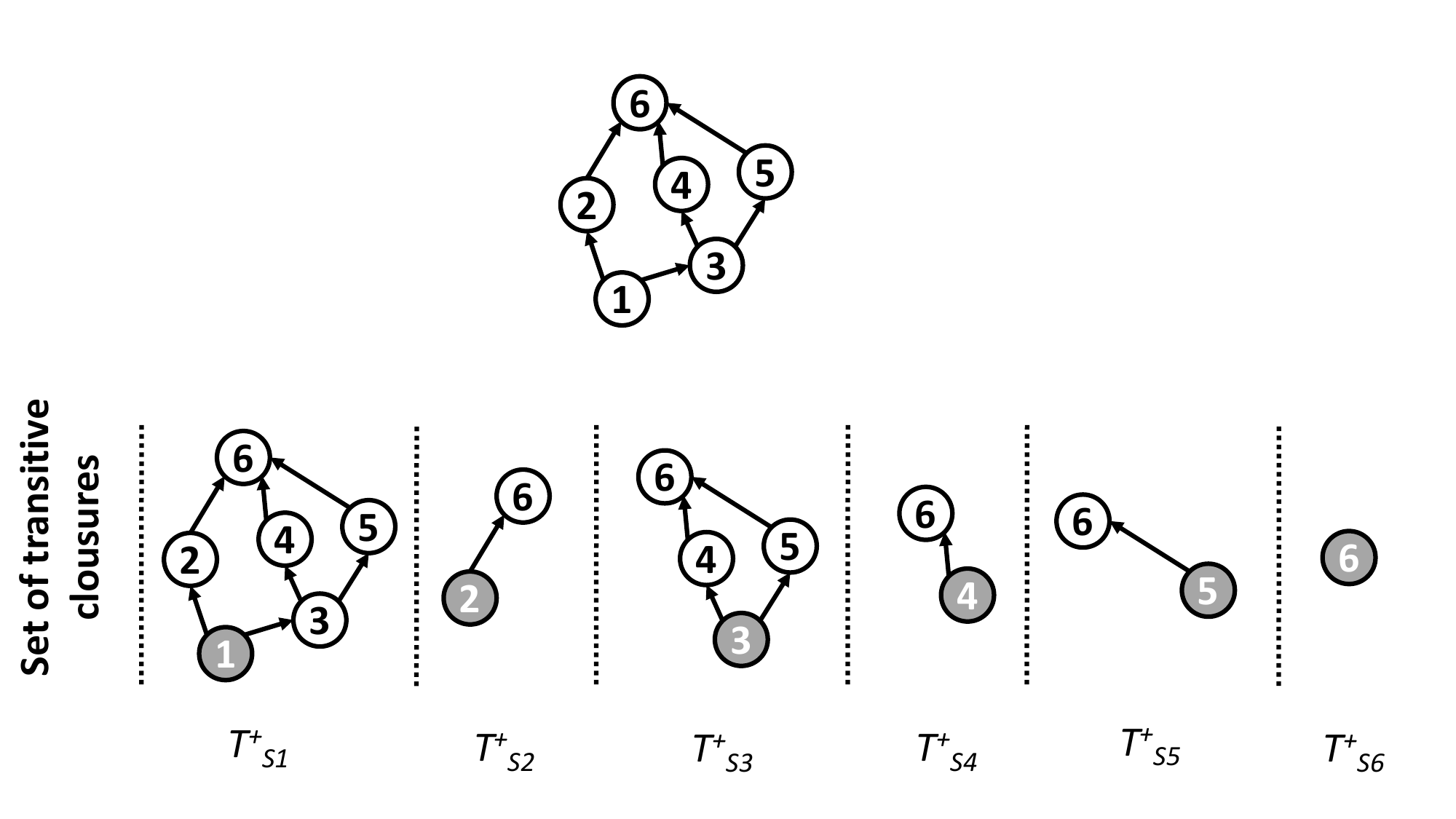}
	\caption{Example of transitive closures for each service of an application.}
	\label{servicesappsubset}
\end{figure}

For an easier explanation of our algorithm in Sect.~\ref{optimizationmodel}, we also consider the transitive closure of a service represented as $T^{+}_{S_x}$. In graph theory, the transitive closure of a node is the set of nodes than can be achieve from that node, i.e., there is at least one path between both nodes~\citep{MUNRO197156}. In our particular case, the transitive closure of a service represents all the services that need to be executed when the service is requested. Figure~\ref{servicesappsubset} shows an example of an application and each of the transitive closures, $T^{+}_{S_x}$, obtained for each service. 




The clients request the applications in the cloud provider through a set of interconnected network devices, $D_i$. These network devices have processing capacities, and they are able to allocate service modules to reduce the network latency or hop distance between clients and services. The physical interconnections of the devices create a graph structure where the nodes are the devices and the edges are the direct network links between devices.

For a clearer explanation of the algorithm in Sect.~\ref{optimizationmodel}, we also defined $SP_{S_x}^{cloud}$ as the shortest path between a device and the cloud provider, that is a path of ordered devices whose sum of the network distances of its constituent network links is the minimum for all the paths between the device and the cloud provider. Additionally, we defined the father of a device, $father(D_1)$, as the first device in the shortest path to the cloud provider.





Several instances, $S^y_x$, of the same service, $S_x$, can be allocated across the system, i.e., the services can be horizontally scaled and clients, or other services, can request any of these instances. The allocation function is defined as a many-to-many relationship $alloc: \{S_x\} \rightarrow \{D_i\}$, if we refer to the services, or as a many-to-one relationship $alloc: \{S^y_x\} \rightarrow \{D_i$\}, if we refer to the instances. Similarly, the many-to-many relationship that represents the services allocated in a device is defined as $alloc: \{D_i\} \rightarrow \{S_x\}$.


The clients need to be connected to the network to request the service modules. These clients are connected to one and only one leaf device, but several clients can be connected in the same leaf device. These clients are mobile devices, sensors, actuators, or others. This connection is modeled through a many-to-one relationship $conn: \{C_n\}  \rightarrow \{D_i\}$. Additionally, each client $C_n$ in the system is characterized by its service request rate, $\lambda^{C_n}_{S_x}$. Consequently, each device $D_i$ in the system can be also characterized by the request rates that it receives for each service, $\lambda^{D_i}_{S_x}$, and calculated as the summation of request rates of the clients whose shortest paths to the cloud providers ($SP_{C_n}^{cloud}$) include that device $D_i$:


\begin{equation}
\lambda^{D_i}_{S_x} = \sum^{C_n} \lambda^{C_{n}}_{S_x}\quad \forall\ C_n\ |\ D_i\ \in\  SP_{C_n}^{cloud}
\end{equation}

The  devices are characterized by their resource capacities. These resources are defined as a set of n-values, one for each resource element, such as processor, memory, or storage. We represent the capacity of a device $D_i$ as a n-tuple $R^{cap}_{D_i} = < r_0, r_1, ... r_{n-1} >$. For the sake of simplicity, we have only considered one resource in this present study, the computational capacity of a device. Thus, the resource capacity is defined as $R^{cap}_{D_i} = < r_{cpu}>$. Services allocated in a device generate a resource consumption that can be defined as a tuple $R^{con}_{S_x} = < r_0, r_1, ... r_{n-1} >$ where the consumption of each individual resource element is indicated. As we have mentioned before, we only consider the CPU consumption in this study, so $R^{con}_{S_x} = < r_{cpu} >$. The cloud provider is consider as an special device where resources are unlimited as they can be scaled horizontally as much as it is necessary. The cloud resource capacity is defined as $R_{cloud} = < \infty >$. The total resource usage of a device, $R^u_{D_i}$, can be consequently calculated as the sum of the resource consumptions of all its allocated services multiplied by the request rate of each services:

\begin{equation}\label{eqRusage}
R^u_{D_i} = \sum^{S_x} R^{con}_{S_x} \times \lambda^{D_i}_{S_x}\quad \forall\ S_x\ |\ S_x\ \in\ alloc(D_i)
\end{equation}

Table~\ref{systemmodel} summarizes the list of the variables defined in the system model, that are used in the following sections of the article.

\begin{table}%
	\caption{Summary of the variables of the system model.}%
	\label{systemmodel}
			\begin{tabular}{lp{0.8\columnwidth}}
		\hline\noalign{\smallskip}
				\textbf{Variable} & \textbf{Description}  \\
		\noalign{\smallskip}\hline\noalign{\smallskip}
				$C_n$ &  A client in the system\\
				$S_x$ &  A service in the system\\
				$S^y_x$ &  An instance of a service $S_x$ \\
				$T^{+}_{S_x}$ & The set of services executed when $S_x$ is requested\\
				$cons(S_x)$ &  Function that returns the list of services that are requested by a given service $S_x$\\
				$D_i$ &  A fog device in the system\\
				$D_{cloud}$ & Identification of the cloud provider\\
				$SP_{S_x}^{cloud}$ & The shortest path between $S_X$ and the cloud provider\\
				$\mathit{father}(D_x)$ & Function that returns, the first element in the shortest path between  $S_x$ and the cloud provider\\
				$alloc(S_x)$ & Function that returns the set of devices where a given service $S_x$ is allocated\\
				$alloc(S^y_x)$ & Function that returns the device where a given service instance $S^y_x$ is allocated\\
				$alloc(D_i)$ & Function that returns the set of services that a given device $D_i$ allocates\\
				$R^{cap}_{D_i}$ & Tuple for the resource capacities of a device $D_i$\\
				$R^{con}_{S_x}$ & Tuple for the resource consumption required by a service $S_x$\\
				$R^u_{D_i}$ & Tuple for the total resource usage in a device  $D_i$\\
				$\lambda^{C_n}_{S_x}$ & The request rate generated by a client $C_n$ over a service $S_x$\\
				$\lambda^{D_i}_{S_x}$ & The request rate that arrives to a device $D_i$ for the service $S_x$\\
				$conn(D_i)$ & Function that returns the list of clients connected to a given leaf device $D_i$\\
				$hop(D_i,D_{i'})$ & Hop count, number of devices, between $D_i$ and $D_{i'}$\\
				
		\noalign{\smallskip}\hline
			\end{tabular}
\end{table}

\subsection{Optimization Model}
\label{optimizationmodel}

The optimization algorithm is based on the idea of placing the most popular services closer to the client as it has been done traditionally in other architecture, as for example, in content delivery networks with the most popular contents \citep{5461964,1250586}) or in web caching \citep{GUERRERO20132970}. We use the service request rate to measure the most popular services.

The algorithm analyzes the request rates of each service in each device, and take local decisions by migrating the less requested services in the device to upper devices ---any device in the shortest path between the current device and the cloud provider--. This decision is based on the idea explained in Sect.~\ref{architectureproposal} and Figure~\ref{shortestPath}, that considers that once that a service is migrated from the desired device, it is better to bring it closer to the cloud provider, i.e., a device in the shortest path.

Additionally, if a service is migrated, all its interoperated services, that are allocated in the same device, are also migrated. This is based on the idea of the device loops also explained in Sect.~\ref{architectureproposal} and in Figure~\ref{deviceLoop}.

Algorithm~\ref{placement} shows the pseudo code of our optimization policy. The SAR is only considered when the candidate service is not already allocated in the device (Line~\ref{alreadyallocated}). The service is directly allocated when the device is the cloud provider (Line~\ref{cloudalloc}) or it has enough free resources (Line~\ref{directallocation}). Additionally, the service allocation can be done only if the total capacity of the device is enough to satisfy the requirements of the service (Line~\ref{checkdevcap}), on the contrary, the SAR is shifted to the upper father device (Line~\ref{shifted2fatCOSnotcapacity}).

If those previous conditions are met, it is necessary to deallocate other services from the device. Since our policy migrates all the interoperated services, instead of a single service, the set of candidates for the migration, $\mathbb{M}_{D_i}$, is formed by all the possible subsets of interoperated services that are currently allocated in the device. Each subset of this candidates' set is obtained from the intersection between the services allocated in the device and each of their transitive closures:

\begin{equation}
\mathbb{M}_{D_i}= \left\{ M_{S_x}^{D_i},\ \forall\ S_x \in alloc(D_i)  \right\}
\end{equation}
where
\begin{equation}
M_{S_x}^{D_i} = \left\{ alloc(D_i)\ \cap\  T^{+}_{S_x} \right\}
\end{equation}
%

For example, if we recover the example in Figure~\ref{servicesappsubset} and we suppose that a given device currently allocates services $S_1$, $S_2$, $S_5$, and $S_6$, the migration candidate set is formed by the following service subsets: $\mathbb{M}_{D_i} = \{\ \{S_1,S_2,S_5,S_6\},$ $\ \{S_2,S_6\},$ $\ \{S_5,S_6\},\ \{S_6\}\ \}$.

Our proposal migrates the candidates by ascending order of the request rates. This is done in line~\ref{getminreq} of Algorithm~\ref{placement}, where , $M_{min}$, is the services' subset in $\mathbb{M}_{D_i}$ with the smallest request rate, and $\Lambda^{{M}_{min}}$ this smallest request rate. Thus, the sets in $\mathbb{M}_{D_i}$ are sequentially selected and deallocated from the less requested one to the most popular, until the freed resources are enough to allocate the candidate service in the device (Line~\ref{whilenotenoughresoruces}) or until the remaining candidates have higher request rates (Line~\ref{nomorelowerreqservices}).

The deallocation is not done straightforward, because the algorithm needs to guarantee that the sets of services with lower rates release enough resources to satisfy the requirements of the candidate service. Thus, a list of pre-released services, $\mathbb{M}_{deallocate}$, is created (Line~\ref{deallocatelistcreation}) and they are finally deallocated only if the freed resources are enough (Lines~\ref{bdeallocatereleasedservices}--\ref{edeallocatereleasedservices}).

The request rate of a services' subset, $\Lambda^{M_{S_x}^{D_i}}$ , is calculated just by the summation of the single request rates, for that fog device, of each service in the $M_{S_x}^{D_i}$:
\begin{equation}
\Lambda^{M_{S_x}^{D_i}} = \sum^{\forall\ S_x \in M_{S_x}^{D_i}} \lambda^{D_i}_{S_x}
\end{equation}

The allocation process starts with each new client connection to a leaf device. The Service Popularity Monitor (SPM) creates a SAR each time that a new client is detected. This detection is performed by analyzing the services requests that are received from the client.


We consider that the hop count is a good indicator to measure the service proximity. The hop count is the number of devices that a request of a client pass through to achieve the device where the requested service is placed. Our algorithm's optimization objective is to reduce the hop count, but favoring the more popular services. We consider the weighted average hop count as an indicator to measure the proximity of the most popular services. We define it as the average of the distance between each service and the clients weighted with their popularity, in terms of request rate:

\begin{equation}\label{eqhopcount}
Weighted\ Average\ Hop\ Count = \sum^{D_i,S^y_x} \frac{\lambda^{D_i}_{S_x}}{\sum^{D_{i'},S^y_{x'}} \lambda^{D_{i'}}_{S_{x'}}} \times hop(C_n,D_i)\ 
\end{equation}

%

where $hop(C_n,D_i)$ is the number of fog devices between the clients and the device that allocates a given instance of a service. A value of 1.0 for the hop count means that all the services are placed in the gateways or leaf devices. The maximum value for the hop count is the number of levels in the network.

The hypothesis of our second research question is that performance metrics, such as network usage or service latency, are improved when the average hop count is reduced.

\IncMargin{1em}

\begin{algorithm}[!t]
	
	\DontPrintSemicolon
	\LinesNumbered
	
	\scriptsize
	\KwIn{$S^y_x$, $D_i$}
	
	\If(\label{alreadyallocated}){$D_i \notin alloc(S_x)$}{ 
		\tcc*{service not allocated in device} 
		
		\If{$D_i$ $=$ $D_{cloud}$}{
			\tcc*{device is cloud provider}
			$alloc(S^y_x) \leftarrow D_{cloud}$ \label{cloudalloc}\;
		}\Else{

			\If{$R^{con}_{S_x} \times \lambda^{D_i}_{S_x}$ $<$ $R^{cap}_{D_i} - R^u_{D_i}$}{
				\tcc*{service demanded resources < device available resources}
				$alloc(S^y_x) \leftarrow D_i $\tcc*{device allocates the service}\label{directallocation}\;
			}\Else{
				$D_{father} \leftarrow father(D_i)$\;
				\If(\label{checkdevcap}\tcc*{service resources $<$ device resources}){$R^{con}_{S_x} \times \lambda^{D_i}_{S_x}$ $<$ $R^{cap}_{D_i}$}{
					Placement($S^y_x$, $D_{father}$) \label{shifted2fatCOSnotcapacity}\tcc*{candidate service migrated to father device} 
				}\Else{
					
					$R^{toFree} \leftarrow  R^{con}_{S_x} \times \lambda^{D_i}_{S_x}$ - $(R^{cap}_{D_i} - R^u_{D_i})$\tcc*{service demanded resources - device available resources}
					$\mathbb{M}_{deallocate} \leftarrow \emptyset$\;
					$\mathbb{S}_{allocated} \leftarrow  alloc(D_i)$\;
					$\lambda^{D_i} \leftarrow calculateM_{D_i}RequestRates(\mathbb{M}_{D_i})$\;
					$\mathbb{M}_{ordered} \leftarrow orderAscBy(\mathbb{M}_{D_i},\lambda^{D_i})$\;
					\tcc*{calculate request rate of each subset and order them}
					
					\While( \label{whilenotenoughresoruces}){$M_{min},\Lambda_{min} \leftarrow \mathbb{M}_{ordered}.next(),$}{\label{getminreq}

						\If(\tcc*{candidate service has higher request rate}){$\lambda^{D_i}_{S_x}$ $>$ $\Lambda_{min} $}{
							
							$ \mathbb{M}_{deallocate} \leftarrow \mathbb{M}_{deallocate} \cup \mathbb{M}_{min} $ \label{deallocatelistcreation} \tcc*{add service set to released list} 
							$ \mathbb{S}_{allocated} \leftarrow \mathbb{S}_{allocated} - \mathbb{M}_{min} $\tcc*{remove released services}

							\For(\tcc*{released services from device}){$S^y_{x'}, \forall\ S^y_{x'} \in \mathbb{M}_{min}$}{	
								
								$R^{toFree} \leftarrow R^{toFree}$ - $R^{con}_{S_{x'}} \times \lambda^{D_i}_{S_{x'}}$\;
								
							}
						}\Else{
							
							break  \label{nomorelowerreqservices} \;
							
						}

					}

					\If(\label{bdeallocatereleasedservices}){$R^{toFree}$ $<=$ 0.0} {
						\tcc*{enough resources once lower request rate services released}
						\For{$S^y_{x'}, \forall\ S^y_{x'} \in \mathbb{S}_{deallocate}$}{
							Placement($S^y_{x'}$, $D_{father}$)\tcc*{released services migrate to father device}
							\label{edeallocatereleasedservices} }
					}\Else{
						
						Placement($S^y_x$, $D_{father}$) \tcc*{candidate service migrated to father device}
						
					}

				}
			}
			
		}

	}
	
	\caption{Algorithm for module placement in devices}
	\label{placement}
\end{algorithm}

%

%
%
%
%
%

\section{Evaluation}

The evaluation of our proposal was done by simulating a microservice-based application in the iFogSim simulator~\citep{SPE:SPE2509}. The simulator's service placement policy was modified by extending the class ModulePlacement. The results of our algorithm were compared with the ones obtained with the simulator's built-in placement policy (Edgewards), since it was, to the best of our knowledge, the only previous decentralized service placement policy.



Several scenario settings were considered by modifying the number of clients, the number of applications and the number of fog devices. Table~\ref{experimentsetting} includes a summary of the simulation settings.

The experiments were characterized with the same configuration parameters than in the study of the simulator's developers~\citep{SPE:SPE2509}. Those experiments considered a tree-based network topology where the number of devices was varied in two dimensions, by ranging the number of levels in the tree, and by ranging the number of children of each fog device. Although our model allows to model the fog architecture as a graph network, the iFogSim only allows to define the architecture like a tree\footnote{The devices of the iFogSim are related with a list of children identificators and just one father, as it can be seen in lines 59 and 68 of class \url{https://github.com/Cloudslab/iFogSim/blob/master/src/org/fog/entities/FogDevice.java}}.

The resource capacities for the cloud were defined high enough to behave as a device with unlimited resources. The memory and the bandwidth  of the fog devices were also high enough to be able to allocated as many services as necessary, keeping the computational capacity the only resource limitation in the service placement, since we only considered CPU resources (Sect.~\ref{optimizationmodel}).

\begin{figure}
	\includegraphics[width=240pt]{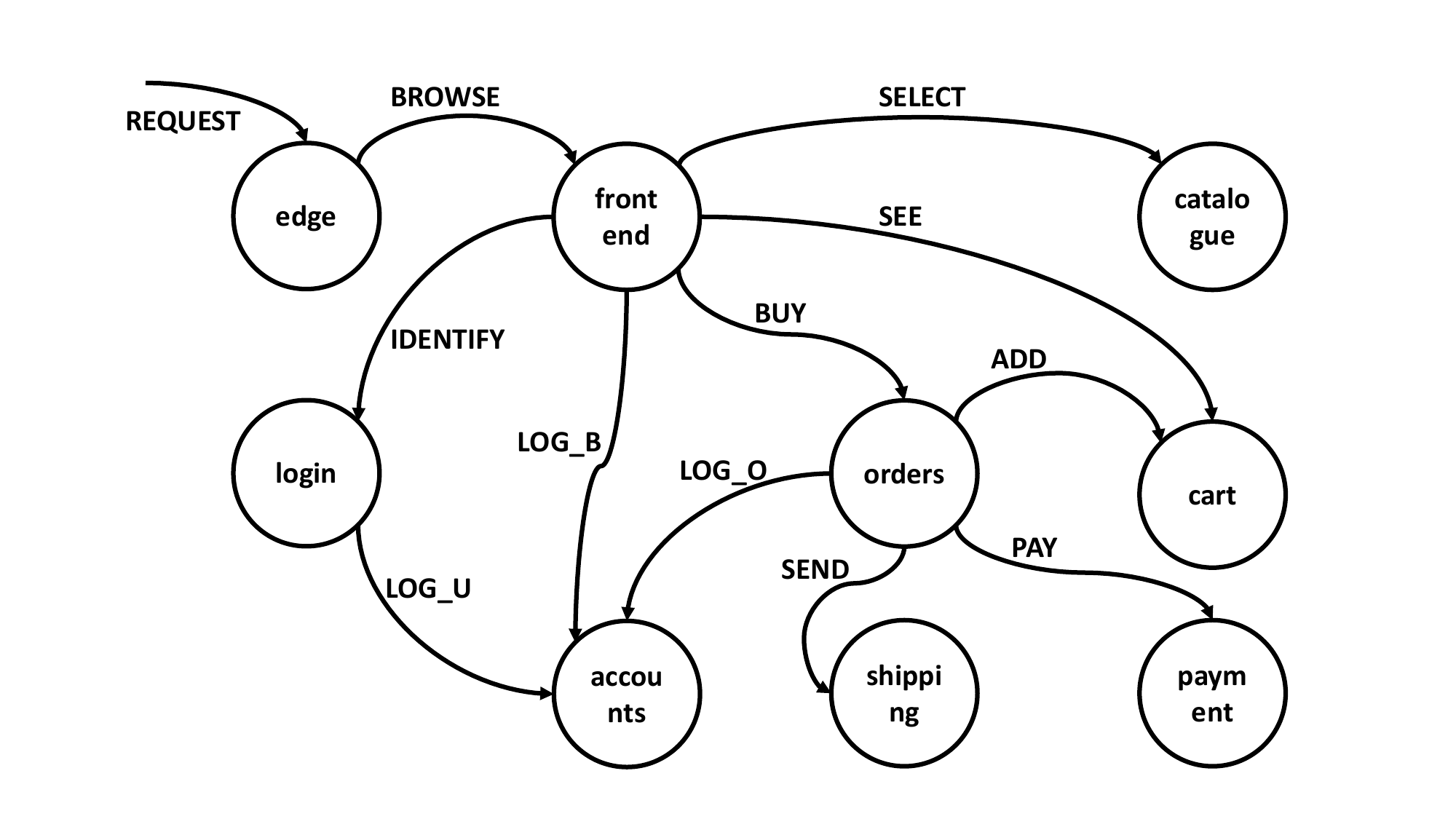}
	\caption{Services, interoperability and application edges settings for Sock shop demo application.}
	\label{appDefinition}
\end{figure}

A microservice-based application were modeled in the simulator as this type of development pattern is also common in IoT applications~\citep{Vogler:2016:SFP:2909066.2850416,7300793,Saurez2016}. We modeled the \textit{Socks Shop}~\citep{weaveworks2016shocksshop} applications, a microservice-based demo that was developed to test the benefits of deploying applications in a container platform. Its characterization and modeling parameters were obtained from previous research works~\citep{Guerrero2017}. Figure~\ref{appDefinition} shows the application graph. The number of applications was varied by replicating the same application several times, but considering different request rate for each replica.

\begin{table}%
	\caption{Summary of experimental settings.}%
	\label{experimentsetting}

			\begin{tabular}{lllr}
				\hline\noalign{\smallskip}
				\textbf{Element} & \textbf{Parameter} & \textbf{Units} & \textbf{Value}  \\
				\noalign{\smallskip}\hline\noalign{\smallskip}
				Cloud & CPU & MIPS & 4480000\\
				& RAM &  MB& 4000000\\
				& BandWith (Up/Down) & bytes/ms & 20000+20000\\
				& Link latency & ms & 100\\
				Fog devices & CPU & MIPS & 2800\\
				& RAM &  MB& 4000\\
				& BandWith (Up/Down) & bytes/ms & 20000+20000\\
				& Link latency & ms & 2\\
				Applications & \# applications &--&[1, 2, 3, 4, 5]\\
				Services & RAM & MB & 1\\
				Service edges & CPU & Instr. $\times\ 10^6$ & 1000\\
				& BandWith & bytes & 10\\
				& Selectivity & Fractional& 1\\
				Users & Request rate & req/ms&[$\frac{1}{10}$, $\frac{1}{20}$, $\frac{1}{25}$, $\frac{1}{30}$, $\frac{1}{35}$]\\
				& \# user per gateway &--&[1, 2, 3, 4, 5]\\
				
				Network topology & \# levels &--&[1, 2, 3, 4, 5]\\
				& \# child devices &--&[1, 2, 3, 4, 5]\\

	\noalign{\smallskip}\hline
			\end{tabular}

\end{table}

\section{Results}

Our results are presented in terms of average hop count, Equation~\ref{eqhopcount}, network usage, and service latency. The equations for the calculation of the two latter metrics, Equations~\ref{eqnetusage} and~\ref{eqservicelatency}, were obtained from the analysis of the source code of the iFogSim simulator\footnote{\label{note1} Obtained from the analysis of the source code available in \url{https://github.com/harshitgupta1337/fogsim}}.


\begin{equation}\label{eqnetusage}
Network\ Usage =\frac{ \sum^{Req_u(D_i,D_{i'})} (T^{lat}_{D_i,D_{i'}} \times NetSize_{Req_u(D_i,D_{i'})}) } { Simulation\ Time }
\end{equation}

where $T^{lat}_{D_i,D_{i'}}$ is the network latency between the devices that are the origin and the target of the request, and $NetSize_{Req_u(D_i,D_{i'})}$ is the total size of the request sent by the network. The network usage is calculated as the sum of the network usages generated by each request $Req_u(D_i,D_{i'})$ sent during the total simulation time.

The service latency is measured in the simulator as the average time to execute a path of interoperated services, called application edge loop. This service latency is calculated as the time between the point in time the request for the first service in the path arrives, $t_{S_{first}}$, and the point in time the last service in the path ends its execution, $t_{S_{end}}$:

\begin{equation}\label{eqservicelatency}
Service\ Latency =\frac{\sum^{Req_u} (t_{S_{end}}-t_{S_{first}})}{|Req_u|}\quad \forall\ Req_u \in Loop(S_{first},S_{end})
\end{equation}

In the following figures, the results obtained with our algorithm are labeled as \textit{Pop} and the ones for the Edgewards policy of the iFogSim are labeled as \textit{Edge}. 

\begin{figure}
	\includegraphics[width=0.45\textwidth]{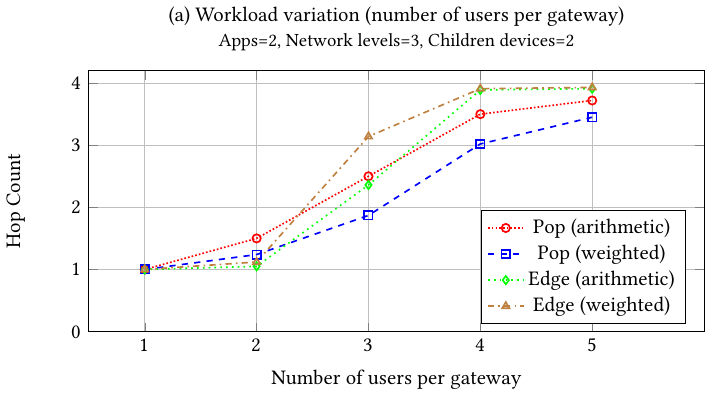}
	\includegraphics[width=0.45\textwidth]{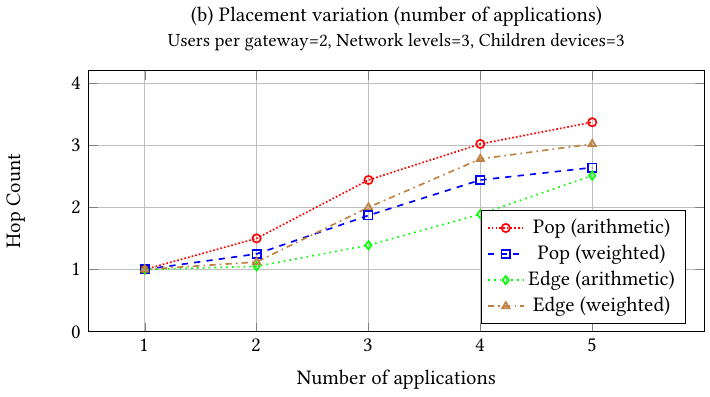}\\
	\includegraphics[width=0.45\textwidth]{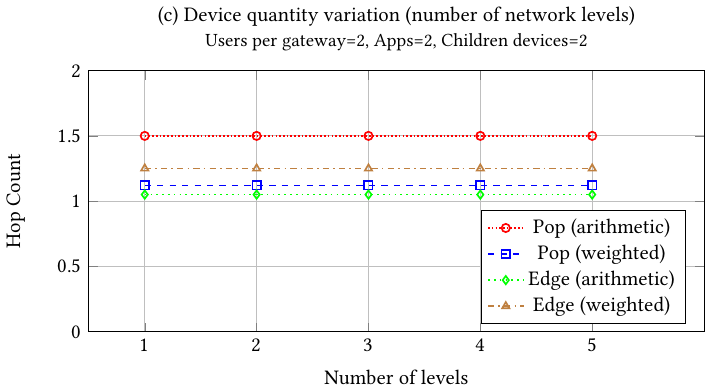}
	\includegraphics[width=0.45\textwidth]{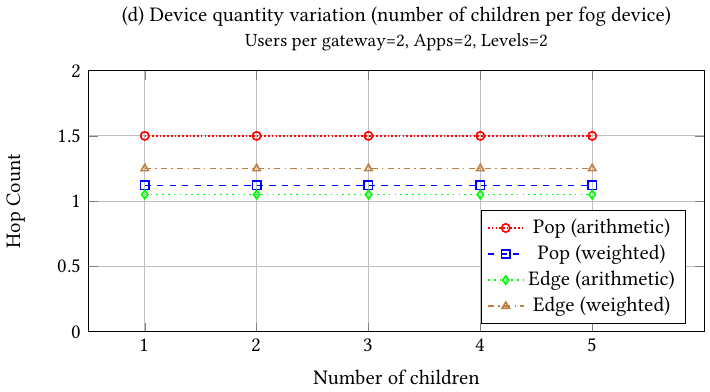}
	\caption{Results for hop count.}
	\label{fighopcount}
\end{figure}
\begin{figure}
	\includegraphics[width=0.45\textwidth]{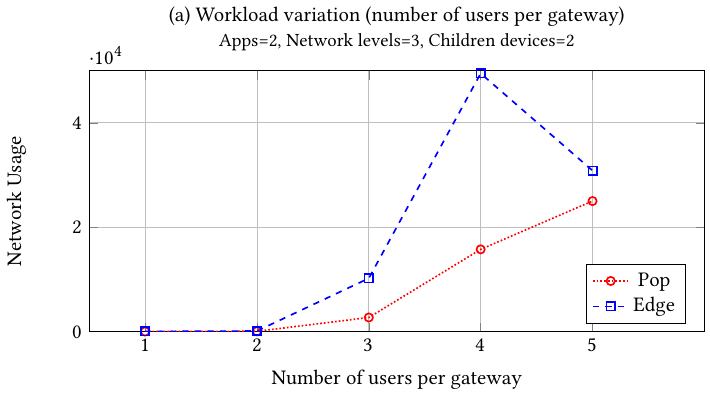}
	\includegraphics[width=0.45\textwidth]{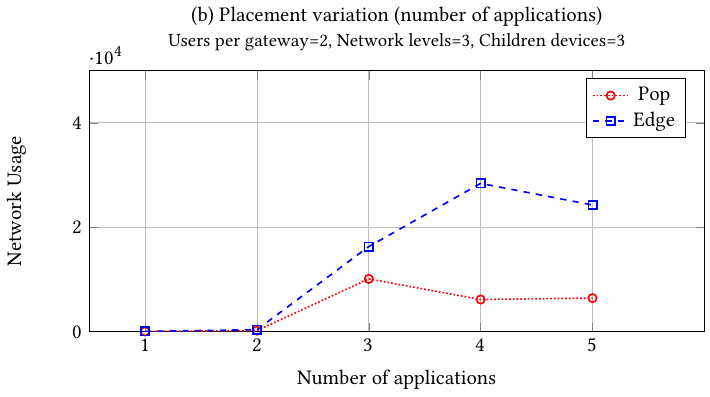}\\
	\includegraphics[width=0.45\textwidth]{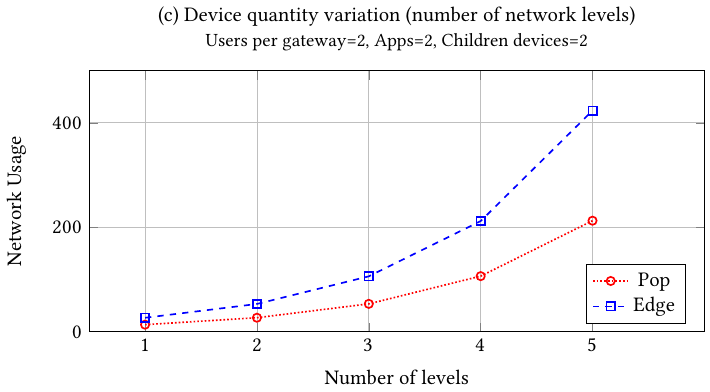}
	\includegraphics[width=0.45\textwidth]{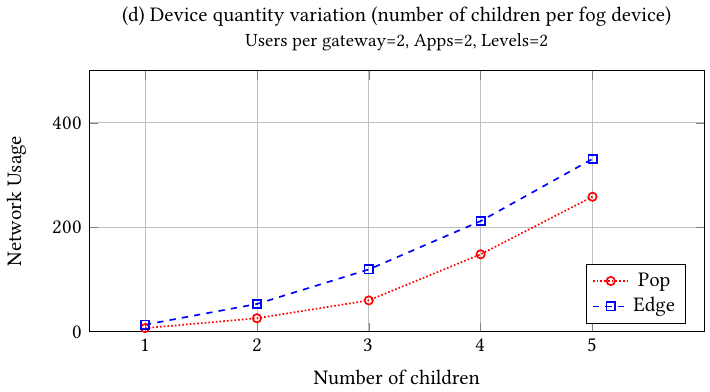}
	\caption{Results for network usage.}
	\label{fignetworkusage}
\end{figure}
\begin{figure}
	\includegraphics[width=0.45\textwidth]{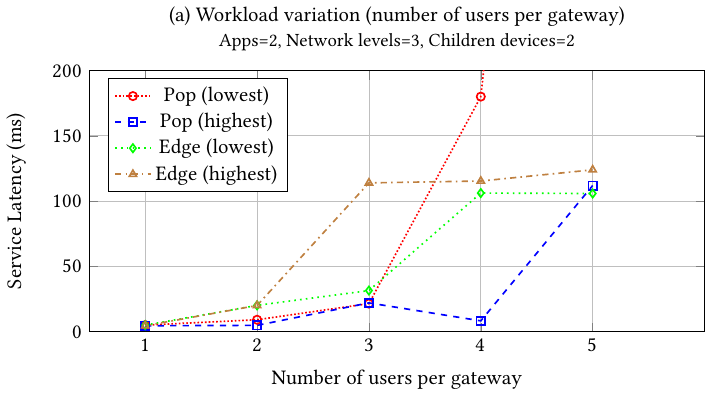}
	\includegraphics[width=0.45\textwidth]{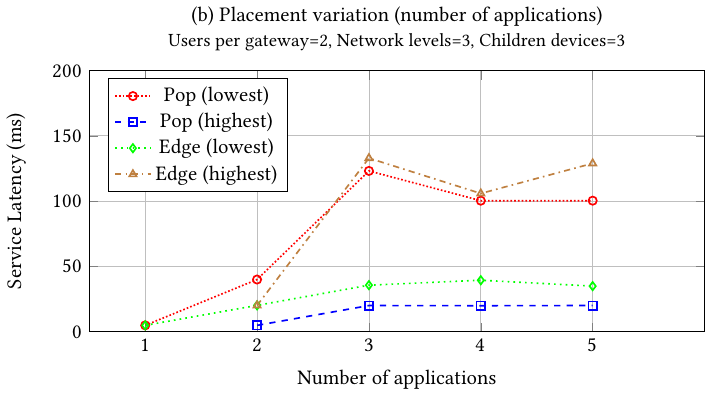}\\
	\includegraphics[width=0.45\textwidth]{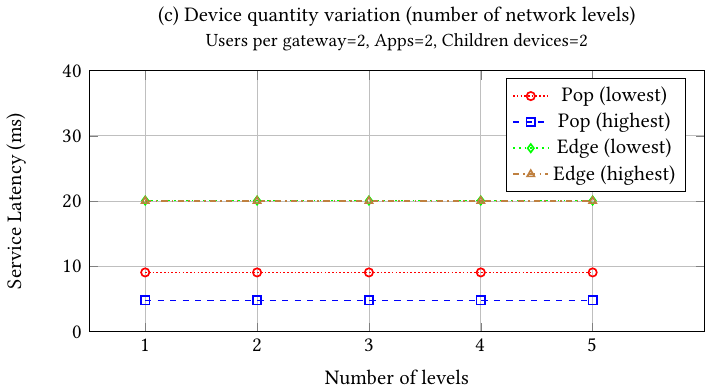}
	\includegraphics[width=0.45\textwidth]{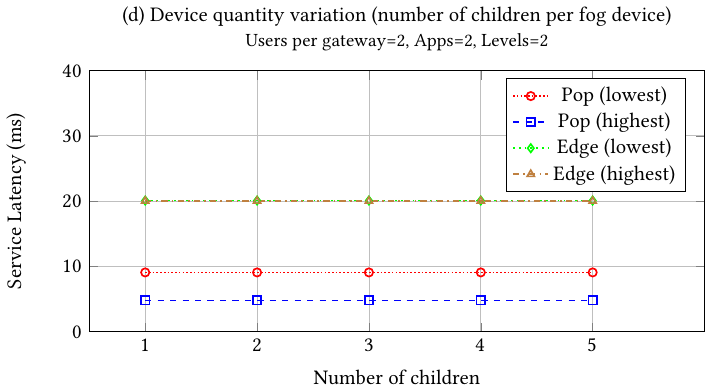}
	\caption{Results for service latency for the service loop \textit{edge, frontend, orders, accounts} of the applications with the highest and the lowest request rates.}
	\label{figservicelatency}
\end{figure}
\begin{figure}
	\includegraphics[width=0.45\textwidth]{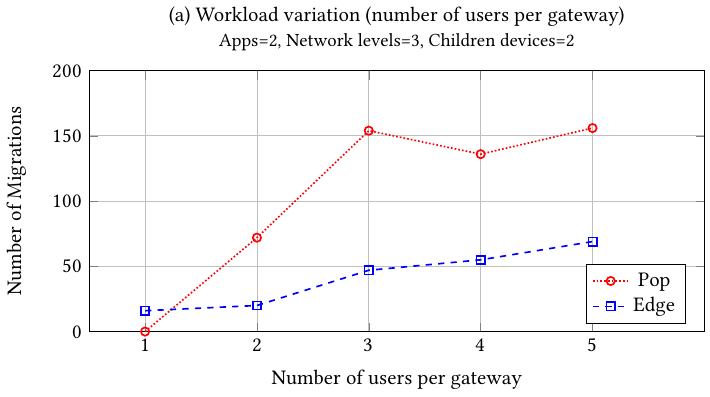}
	\includegraphics[width=0.45\textwidth]{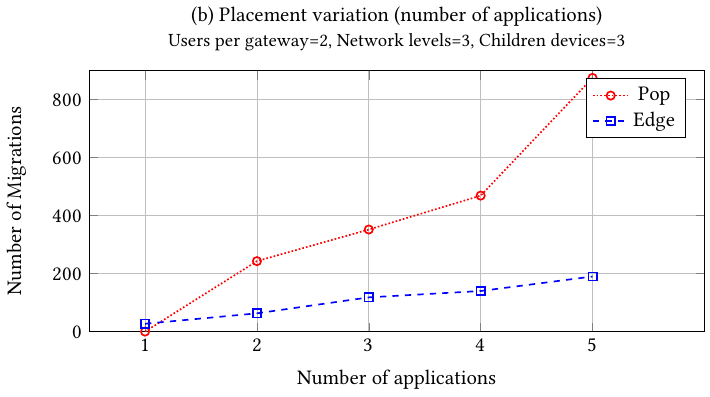}\\
	\includegraphics[width=0.45\textwidth]{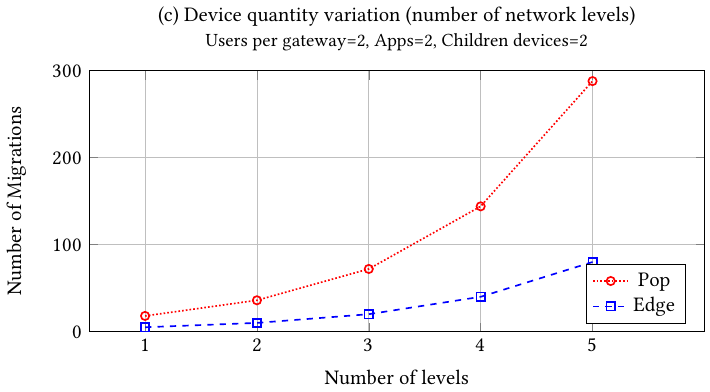}
	\includegraphics[width=0.45\textwidth]{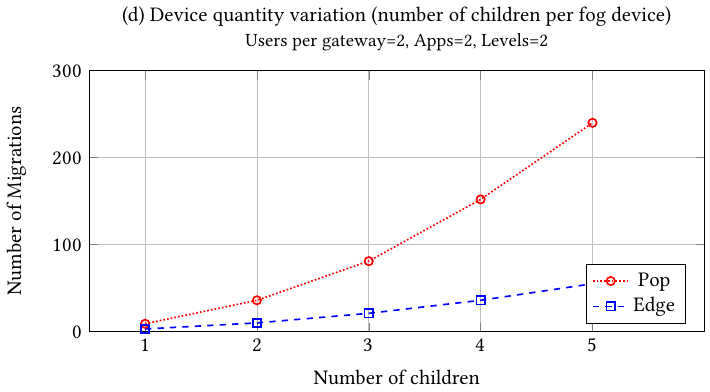}
	\caption{Results for the number of migrations.}
	\label{figmigrationcount}
\end{figure}

The first set of figures (Figures~\ref{fighopcount}, \ref{fignetworkusage}, \ref{figservicelatency}, and~\ref{figmigrationcount}) includes four subfigures one for each of the size variations in the execution settings: (a) variations in the number of users connected to each gateway to evaluate different levels of workload; (b) variations in the number of applications available in the system to evaluate different number of services to place, i.e., different number of fog devices allocating services; (c) variations in the number of levels of fog devices between the users and the cloud provider to evaluate the influence of the path length; and finally, (d) variations in the number of children devices that each fog device is connected to with the objective of evaluating the influence of the number of devices.

Figure~\ref{fighopcount} shows the results for the hop count. It plots the weighted average ---calculated with Equation~\ref{eqhopcount} and labeled as \textit{weighted}--- and the arithmetic mean ---labeled as \textit{arithmetic}---. The arithmetic mean represents the overall proximity between the users and all the services. On the contrary, the weighted mean represents how closer the most requested services are to the clients. 

Figure~\ref{fignetworkusage} shows the results calculated by the simulator with respect to the network usages.  Figure~\ref{figservicelatency} shows the latency results for  a representative service loop of the application.  We chose the loop of \textit{edge}, \textit{frontend}, \textit{orders}, and \textit{accounts} services as it includes the service with the higher rate, the \textit{accounts} with 3.0 requests for each request that arrives to the \textit{edge}. The value that the simulator calculates for the loop represents the time between the \textit{edge} server is requested and the \textit{accounts} one finishes. To illustrate the benefits of our algorithm for the services with the highest request, the plots represent the service latency for the loops of the applications with the lowest, labeled as \textit{lowest}, and highest request rate, labeled as \textit{highest}.

Figure~\ref{figmigrationcount} represents the total number of migrations performed during the service placement process. It is important to remember that the applications are defined as stateless services and consequently a migration consist on removing the current service instance and downloading a new instance from the cloud to the new device. 

\begin{figure}
	\includegraphics[width=0.95\textwidth]{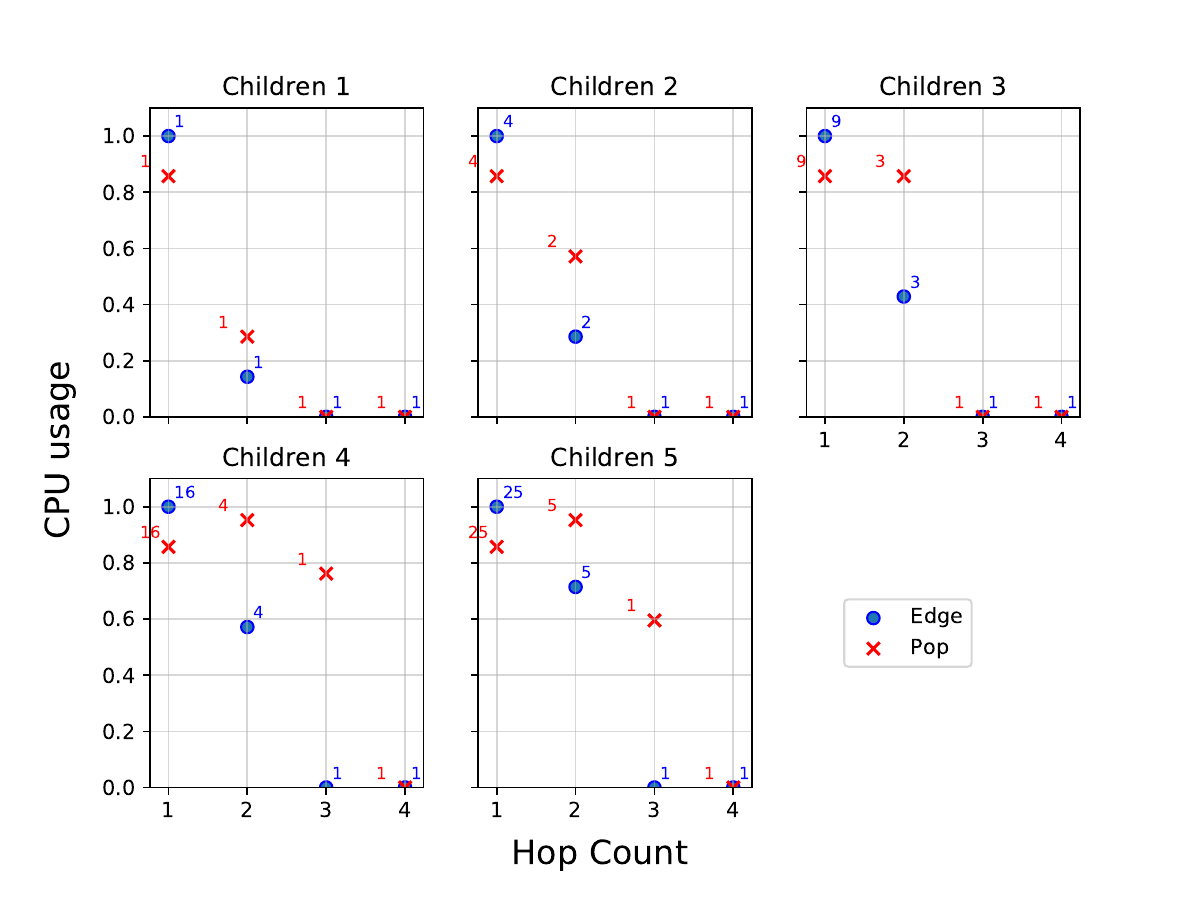}
	\caption{CPU usage of the devices with regard to their topology distribution. Experiment with 2 applications, 2 users, and 2 levels of fog devices.}
	\label{figCPUa2l2u2cX}
\end{figure}

\begin{figure}
	\includegraphics[width=0.95\textwidth]{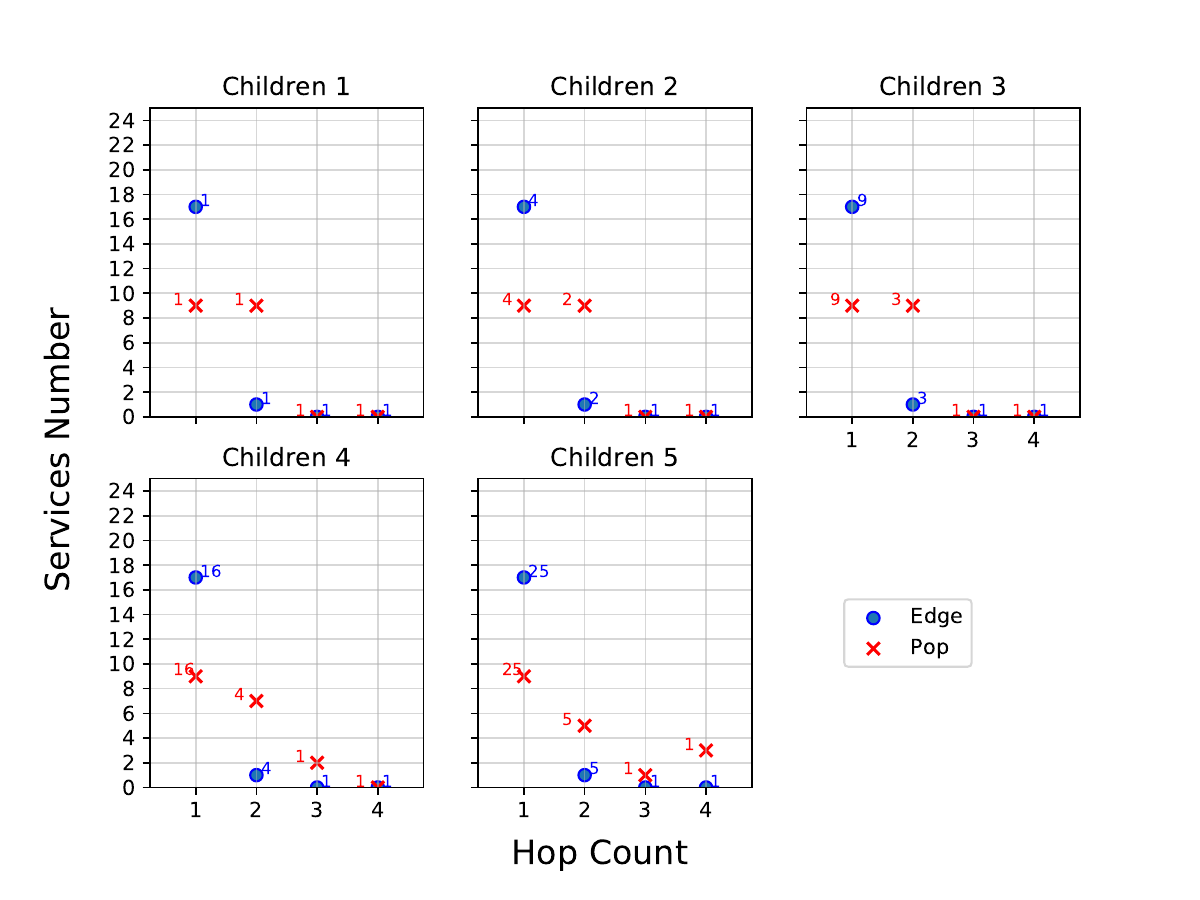}
	\caption{Number of services in the devices with regard to their topology distribution. Experiment with 2 applications, 2 users, and 2 levels of fog devices.}
	\label{figSERa2l2u2cX}
\end{figure}

\begin{figure}
	\includegraphics[width=0.95\textwidth]{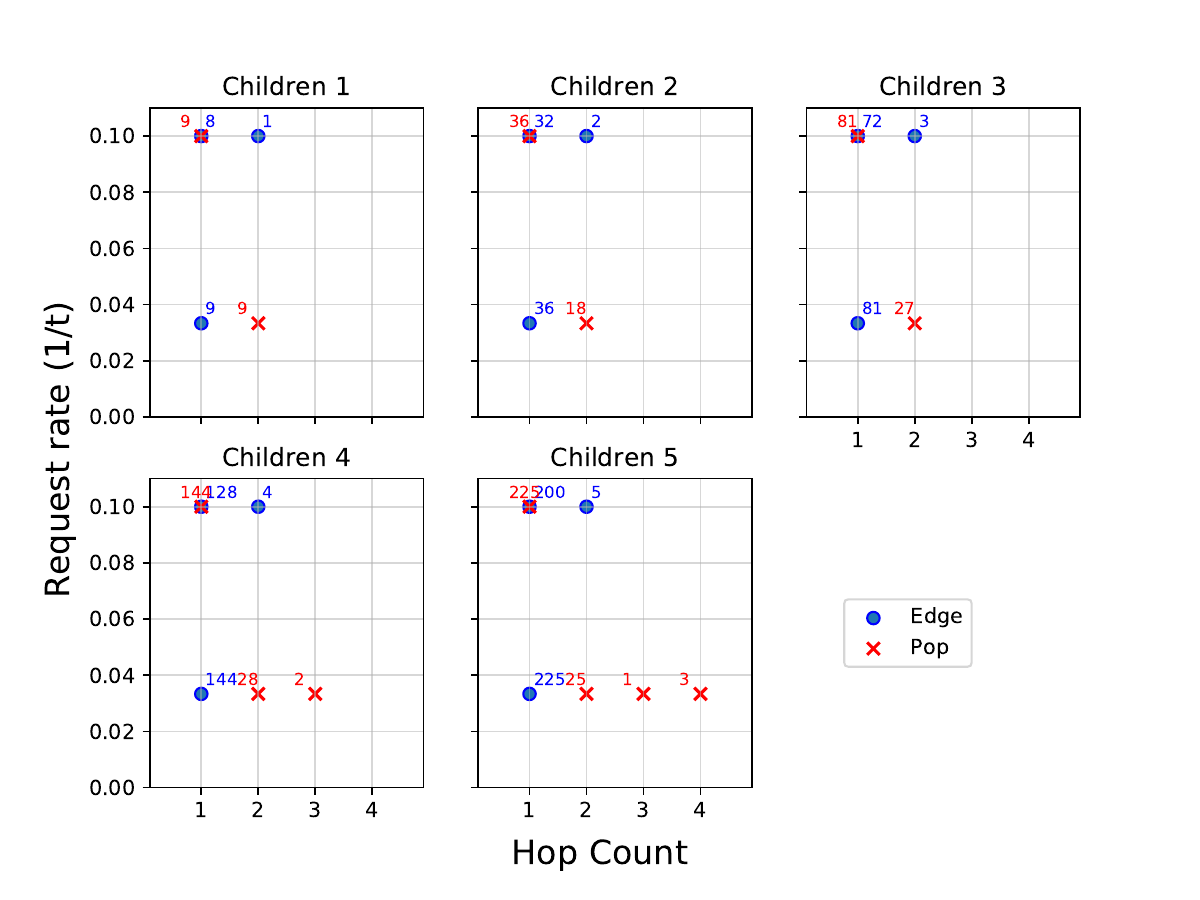}
	\caption{Distribution of the services across the device topology with regard to their request rate. Experiment with 2 applications, 2 users, and 2 levels of fog devices.}
	\label{figRATa2l2u2cX}
\end{figure}

The last set of figures (Figures~\ref{figCPUa2l2u2cX}, \ref{figSERa2l2u2cX}, and~\ref{figRATa2l2u2cX}) represent the relationship of the distance between the IoT devices and the service placement distribution, measured in terms of the hop count, in regard with the popularity of the services (request rate), the number of services in a device and the CPU usage. Those distributions are represented in one independent plot for each experiment size. Consequently, they are grouped in sets of five plots, one for each single size of a variation set.  Thus, there are four sets of five plots of figures for each of the three cases (request rate, number of services, and CPU usage). We present only some representative cases of those plots, particularly, we include the cases for the children variations.

Figure~\ref{figCPUa2l2u2cX} shows the CPU usage of the devices by classifying the devices by their distance from the IoT devices (sensors). The distance is measured in terms of hop count (x-axis) and the CPU usage is the rate between consumed and total resources (y-axis). Each point of the plot represents a device with its corresponding usage value. The number of each point indicates the total number of samples (devices) with the same hop count and CPU usage. The figure includes the results for the experiment with a size of 2 applications, 2 users per IoT gateway, 2 fog devices levels and a range of 1 to 5 children devices per device.

Figure~\ref{figSERa2l2u2cX} is very similar to the previous one, with the only difference that it represents the number of services allocated in the devices instead of the CPU usage.

Finally, Figure~\ref{figRATa2l2u2cX} represents how the services are distributed across the devices in the topology (classified by the hop count with the user) with regard to their request rate (y-axis). The request rate is measured in terms of frequency (the inverse of the time unit). Consequently, each point of the plot represent a service with its allocation in the topology (the hop count of the device where it is allocated) and its request rate.

\section{Discussion}

We use the weighted average hop count as an indicator of the proximity between the clients and the most popular services. On the contrary, the arithmetic average hop count is an indicator of the proximity between all the services and the clients, independently of their request ratios. Therefore, the first research question (RQ1) is answered by the analysis of the series labeled as \textit{weighted} in Figure~\ref{fighopcount}. This metric is, in general terms, smaller for the case of our policy than for the Edgewards, obtaining an overall improvement of the 12\%. There are only two cases in which the Edgewards policy shows smaller values (2 users, 3 levels and 2 applications).  On the contrary, Edgewards policy shows smaller values for the arithmetic average hop count. It means that our policy obtains better proximity for the most requested services, at the expense of increasing the overall distance of the services. We finally observe that cases with only 1 user can be managed by placing all the services in the leaf devices (gateways), since the hop count value is 1.0.

The second research question (RQ2) is answered by the analysis of results in Figures~\ref{fignetworkusage} and~\ref{figservicelatency}. Firstly, our policy's network usage is always smaller than the Edgewards. The improvement in this metric is measured with an speedup between 23\% and 362\%, with an mean value of 114\%. Secondly, the analysis of the service latency is separately done for the case of the applications with the highest and the lowest request rates. In general terms, our policy shows an improvement of the service latency for the application with the highest rate, at the expense of degradation for the application with the lowest request rate in some experiments.

The benefit of our policy for the application with the highest request rate is observed in Figure~\ref{figservicelatency} where the series \textit{Pop (highest)} are the ones with the smallest values. Our policy's improvement is measured in a speedup around 300\% and 500\%, obtaining even an speedup of  1300\% for the second last case of the number of users variations.

The degradation of our policy for the less requested application is observed, mainly, for the experiments where the number of applications is increased, i.e. the cases with higher workload in the system. On the contrary, our policy is better even for the low requested services when the workload of the system is low. This is expected since the more application replicas, the more devices allocating services, and consequently, the less requested applications are much further from the clients, and their latencies are increased. Edgeward policy shows improvements ranging from 60\% to 250\% for this second type of applications in the high workload experiments.


From the analysis of the number of migrations required for the deployment of the services (Figure~\ref{figmigrationcount}), it is observed that our solution clearly needs a higher number of migrations (or service deployments from the cloud provider). This generates a higher network usage due to the download of the services from the cloud provider. It is important to highlight that this process is just performed during the deployment of the application. This migration cost is made up with the benefits of our service distribution, except for the cases with high rates of application deployments. 

It is important to highlight that the number of devices does not influence in the latency neither the hop count in the experiments for the variation in the number of levels or children. This is explained because all the service placements are done in devices in the lowest levels, and devices from upper levels are not necessary and, consequently, the service latency is constant as the number of devices is increased. This is also validated with the results of, for example, Figure~\ref{figSERa2l2u2cX}, where the number of allocated services is 0 for the devices with a hop count higher than 2. On the contrary, the network usage is influenced because the number of connections between the devices is varied, and this metric is influenced by the number of those connections.
 
 Figure~\ref{figSERa2l2u2cX} also shows us that our policy migrates more services to upper devices, than the Edgewards algorithm. This is because we do not migrate only one service but also all the consumed services ($M_{S_x}^{D_i}$) to avoid device loops in the service execution flow. By this, the less requested services are placed in the devices with a hop distances of 2, and the most requested one in the devices with hop distance of 1. Consequently, we can also observed how the devices in the lower levels of the topology have smaller usages than in the case of the  Edgewards (Figure~\ref{figCPUa2l2u2cX}). This offers a side effect of a lower saturation of the devices and an evenly distribution of services.
 
 Finally, Figure~\ref{figRATa2l2u2cX} reflects how the services that are placed in lower devices (devices with a hop count of 1) have higher rates in the case of our policy with regard to the Edgewards. Likewise, the devices with a hop count of 2 allocate services with lower request rates.

To sum up, we can conclude that our policy reduces the distance between the clients and the most requested services (RQ1), and the latency of those services and the overall network usage are improved (RQ2). On the contrary, the latency of the less requested application is degraded in experiments with the highest workloads. This is obtained by increasing the number of service migrations.

\section{Conclusions}

We have presented a decentralized algorithm for the Fog Service Placement Problem to optimize the distance between the clients and the most requested services. The algorithm is locally executed in each fog device, by considering only performance and usage data obtained in the device itself.  The decision of the service placement is addressed to migrate services with smaller request rates to upper devices, as new placements are necessary. The interoperability between the services is considered and the migration of a service also involves the migration of all its consumed services.

We have evaluated our placement policy with the iFogSim simulator by modeling a microservice-based application with several experiment sizes. Our results have been compared with the simulator's built-in policy, Edgewards. The results showed that our policy reduces the distance between the clients and the most requested services, measured in terms of the weighted average hop count. Consequently, the network usage and the service latency of those services are improved. These improvements are obtained at the expense of an increment of the latency in the less requested services.

Our proposal shows that decentralized placement optimization in fog computing is able to improve the overall performance of the system. This opens new challenges to apply other optimization techniques to decentralized managers to avoid, between others, scaling problems of the centralized ones.

\bibliographystyle{spbasic}      
\bibliography{sample-bibliography}   


\end{document}